# Elevating the Role of Software as a Product of the Research Enterprise

*Arfon M. Smith <arfon@stsci.edu> Space Telescope Science Institute, Dara Norman <dnorman@noao.edu> NOAO , Kelle Cruz <Kelle.cruz@hunter.cuny.edu> Hunter College-CUNY, Vandana Desai <desai@ipac.caltech.edu> IPAC, Eric Bellm <ecbellm@uw.edu> University of Washington, Britt Lundgren <blundgre@unca.edu> University of North Carolina Ashville, Frossie Economou <frossie@lsst.org> LSST, Brian D. Nord <nord@fnal.gov> Fermilab/University of Chicago, Chad Schafer <schafer@cmu.edu> Carnegie Mellon University, Gautham Narayan <gnarayan@stsci.edu> Space Telescope Science Institute, Joseph Harrington <jh@physics.ucf.edu> University of Central Florida, Erik Tollerud <etollerud@stsci.edu> STScI, Brigitta Sipőcz <brigitta.sipocz@gmail.com> University of Washington, Timothy Pickering <tpickering@mmto.org> MMT Observatory, Molly S. Peeples <molly@stsci.edu> Space Telescope Science Institute/JHU, Bruce Berriman <gbb@ipac.caltech.edu> IPAC, Peter Teuben <teuben@astro.umd.edu> University of Maryland, David Rodriguez <drodriguez@stsci.edu> Space Telescope Science Institute, Andre Gradvohl <gradvohl@ft.unicamp.br> University of Campinas, Lior Shamir <lshamir@ksu.edu> Kansas State University, Alice Allen <aallen@ascl.net> Astrophysics Source Code Library/University of Maryland, Joel R. Brownstein <joelbrownstein@physics.utah.edu> University of Utah, Adam Ginsburg <aginsbur@aoc.nrao.edu> NRAO, Manodeep Sinha <msinha@swin.edu.au> Swinburne University of Technology, Cameron Hummels <chummels@gmail.com> Caltech, Britton Smith <britton.smith@ed.ac.uk> University of Edinburgh, Heloise Stevance <hfstevance@gmail.com > University of Sheffield/University of Auckland, Adrian Price-Whelan <adrn@astro.princeton.edu> Princeton, Brian Cherinka <bcherinka@stsci.edu> Space Telescope Science Institute, Chi-kwan Chan <chanc@email.arizona.edu> Steward Observatory, Jeyhan Kartaltepe <jeyhan@astro.rit.edu> Rochester Institute of Technology, Matthew Turk <mjturk@illinois.edu> School of Information Sciences, Benjamin Weiner <bjw@mmto.org> Steward Observatory, Maryam Modjaz <mmodjaz@nyu.edu> NYU, Robert J. Nemiroff <nemiroff@mtu.edu> Michigan Technological University, Wolfgang Kerzendorf <wkerzendorf@gmail.com> ESO, Iva Laginja <ilaginja@stsci.edu> Space Telescope Science Institute, Chuanfei Dong <dcfy@princeton.edu> Princeton, Bruno Merín <bmerin@sciops.esa.int> ESAC, Jennifer Sobeck <jsobeck@uw.edu> University of Washington, Derek Buzasi <dbuzasi@fgcu.edu> Florida Gulf Coast University, Jacqueline K Faherty <jfaherty17@gmail.com>, American Museum of Natural History,*

*Ivelina Momcheva <imomcheva@stsci.edu> Space Telescope Science Institute, Andrew Connolly <ajc@astro.washington.edu> University of Washington, V. Zach Golkhou <zgolkhou@uw.edu> University of Washington*

Infrastructure Activity, State of the Profession, Other: Workforce

Endorsed by: Luigi Foschini <luigi.foschini@inaf.it> Istituto Nazionale di Astrofisica (INAF), Andrew Wetzel <awetzel@ucdavis.edu> UC Davis, Thomas Aldcroft <taldcroft@cfa.harvard.edu> Harvard Center for Astrophysics

**350 character summary (for the submission form):** Software is a critical part of modern research, and those responsible for its development must be retained in the workforce to maximize the scientific return from petabyte-scale datasets. This white paper will present current challenges and suggest practical solutions for elevating the role of software as a product of the research enterprise.

# 1. Executive Summary and Recommendations

Software is a critical part of modern research, and yet there are insufficient mechanisms in the scholarly ecosystem to acknowledge, cite, and measure the impact of research software. The majority of academic fields rely on a one-dimensional credit model whereby academic articles (and their associated citations) are the dominant factor in the success of a researcher's career. In the petabyte era of astronomical science, citing software and measuring its impact enables academia to retain and reward researchers that make significant software contributions. These highly skilled researchers must be retained to maximize the scientific return from petabyte-scale datasets. Evolving beyond the one-dimensional credit model requires overcoming several key challenges, including the current scholarly ecosystem and scientific culture issues. This white paper will present these challenges and suggest practical solutions for elevating the role of software as a product of the research enterprise.

## Recommendations

1. *Recognize software as part of the career path:* Software should be recognized in hiring and career development as a key product of modern astronomical research. Software outputs should be considered in all aspects of academic performance appraisals, career applications, and promotion and tenure review cases.

2. *Improve adoption of best practices for software citation:* Journals and reviewers should adopt best practices for assuring that software and other science support infrastructure is properly referenced and cited in articles. Referees and other reviewers should be trained to recognize when such acknowledgement is necessary and ask authors to provide that information.

3. *Adopt promotion metrics that acknowledge software and other science support*: Departments and other members of the community should adopt and use suggested metrics for promotion and tenure reviews of those scientists whose work and contributions involve software and science infrastructure.

4. *Explore partnerships to support data science staff:* Universities should explore means of providing support for data-science faculty and staff, perhaps sharing capacity between academic groups or departments internally or partnerships outside the university.

5. *Support long-term technical capacity:* Funding agencies should explore longer-term grants aimed at building and supporting professional (non-student) data science capacity.

6. *Treat software milestones as scientific milestones:* A major (community) software milestone or achievement should be seen as having the same value as a major scientific discovery. For example, key developers of astronomy community software should have the opportunity to give plenary talks in major *astronomy* conferences, not just in, e.g, SciPy conference.

7. *Create community prizes for software contributions:* Professional astronomical societies (and other organizations/agencies that support astronomical research) should create dedicated prizes and allow for software contributions to be recognized as criteria of merit within existing prizes.

## 2. Identifying career paths around scientific software development & science with 'big data'

The key skills necessary for data-intensive scientific research are also highly valued in industry, government, and media/communication sectors. Adequate training in astronomy can serve as a stepping stone to fulfilling careers in a wide variety of fields. Since existing working relationships strengthen any partnership opportunities, astronomers should support and encourage those who transition to jobs in non-academic science. However, we need informed people on both sides of these partnerships. In many cases, challenging and uncertain career paths in astronomy push those skilled in software development towards careers where their contributions are more readily appreciated and recognized. This 'brain drain'[1] siphons away the very researchers most needed to tackle the pressing science questions of the 2020s.

### 2.1 Career paths at universities

In the university context, tenure-track faculty positions remain the gold standard for job stability, compensation, and prestige. However, despite the fundamental role of software in scientific discovery, it remains challenging to receive credit towards tenure and promotion for developing software and data services. Later in this paper, we offer specific recommendations for improving recognition for these contributions.

---
[1] https://jakevdp.github.io/blog/2013/10/26/big-data-brain-drain/

Even with appropriate credit for software contributions, faculty positions will continue to carry expectations of leadership, grant-writing, teaching, mentorship, and service, as is appropriate. Furthermore, driven by ongoing changes in the landscape of higher education, tenure-track hiring continues to flatten. To benefit from the opportunities of large datasets, universities also need the ability to support and retain technically capable faculty and staff who have expertise and long-lasting career paths that typically cannot be matched by graduate students or postdocs. These 'research software engineers'[2] would provide a technical core for data-intensive research groups, just as opto-mechanical and electrical engineers are vital to the success of instrumentation labs.

Stable funding is the most significant need for the success of staff research software engineers (Geiger et al, 2018). A patchwork of 2-3 year soft-money grants is insufficient to retain highly-capable professionals, especially when industry salaries are significantly higher. Universities should explore means of providing internal support for data science staff, perhaps sharing capacity between academic groups or departments.
Long-term vision and leadership in the field are needed to recognize and measure relevant metrics and make them part of advancement/career ladders. Retention of technology staff with deep domain knowledge is essential in promoting efficiency, given that rapid growth in data volume is driving a corresponding growth in the size and complexity of software.

## 2.2 Career paths at science and data centers

At data centers, project data management (DM) teams need to cover a wide range of expertise such as astronomical domain knowledge, strong astronomical data understanding, deep software engineering skills and what is often referred to as 'dev-ops' skills (engineering, deploying and operating production services). Given the broad areas of competency required, a team with a couple of people (or worse, sub-teams) in each area of expertise quickly exceeds the 'optimal team size'. This leads to communication overheads, difficulty forming a common purpose and a loss of agility. This may present itself as over-planning, inability to respond to shifting requirements or technologies, and make-work to compensate for inhomogeneities in the division of labor. A more refined approach would be to assemble a hybrid team. A hybrid team is not only multi-disciplinary but also has generalists that are fluent in several domains.

---

[2] https://researchsoftware.org

By assembling hybrid teams that bring domain speciality and share a common understanding of other areas in the team's competence sphere, it is possible to constrain a team to a manageable size. This avoids over-division of labor and the fractioning of individuals' work assignments. This also taps the ability of multi-disciplinary teams to reach new, overarching insights into their problem space. Developing these hybrid teams includes supporting tech-savvy researchers who have expertise in both the domains of astrophysics and software engineering or other data management skills. Ultimately, supporting these hybrid teams requires investment in job stability. Longer-term grants and funded data mission support aimed at building and supporting abiding, professional (non-student) data science capacity are needed.

## 3. Elevating the role of software as a product of the research enterprise

The ecosystem around publishing scholarly work does not properly capture contributions to scientific discoveries made through tools, services and related research infrastructure. Changes need to be made to account for how modern science is done. Any changes to the publication policies can only be achieved after the board, comprised of professional astronomers, are convinced that software is a cornerstone of modern research. Therefore, it is crucial to educate the larger community on changes needed to support modern recognition standards for software services and then advocate for these changes with the membership and their professional society leadership.

We will need major changes to prevalent social and cultural traditions to achieve, as standard practice, appropriate acknowledgement of people who create and maintain software infrastructure tools. We need academic leadership, professors and other staff in positions of power (e.g., on promotion and tenure review committees, recruitment teams, grant review panels) to value software as an important product and enabler of research. Although change takes time, it is important that we begin making those changes with concrete and practical recommendations that can be incrementally introduced in what become acceptable policies, procedures and communal norms. These recommendations include the identification of metrics that support a proper assessment of the impact of software on achieving scientific results.

Over the last decade, substantial improvements to enable the citation of software and tracking of these citations has been made in astronomy and astrophysics, but additional advancements are needed.

Since the Astro2010 papers by Weiner et al. (2009): "Software release should become an integral part of the publication process." and "The barriers to publication of methods and descriptive papers should be lower.", considerable progress has been made in this area. For example, over the past decade, NASA has improved its software release practices, as can be seen with the development of sites such as http://code.nasa.gov, and has provided funding for the Astrophysics Source Code Library[3] (ASCL) to make NASA-funded open codes used in published research and citations to these software packages more discoverable in ADS via the ASCL. (Allen & Schmidt 2015). The American Astronomical Society (AAS) journals now allow software-only publications on equal footing with more traditional science publications[4] and other major astronomy journals like A&A and PASP do as well. New approaches to publication like the Journal of open source Software[5] (Smith et al. 2018) and the ASCL are now providing alternate ways to publish software that are indexed in Astrophysics Data System (ADS). Software archives like Zenodo[6] now connect with GitHub to make the publication of software via DOI almost frictionless[7]. While there are still challenges in identifying how software citation should work in these areas, tangible progress and recommendations are being made (Smith et al. 2016). The "cultural" elements of ensuring these publications are viewed with the same level of value as other publications may also be improving, although concrete data in this area is lacking. While somewhat less progress has been made in ensuring open software is a truly integral part of publication, the same resources noted above have made it much easier to preserve software long-term. More challenging is preserving the environment software has been run in. While technologies like Docker or virtualization provide a possible path, they have not been adopted widely across the community thus far, and represent a possible major area of development for the 2020s.

3.1 Measuring and citing the impact of software

One key factor for improving the recognition of software within academia is to enable native software citation, that is, make it possible and required for authors to cite the software packages they have used in the process of carrying out their research, and to then count these citations in tools such as the Astrophysics Data System (ADS). Enabling software citation is both a technical challenge and a cultural one: prescriptions for what software should be cited, and when to cite it, have been explored in community-wide efforts at FORCE11[8] (Smith et al. 2016), and follow-on efforts are

---

[3] http://ascl.net
[4] https://journals.aas.org/policy-statement-on-software/
[5] http://joss.theoj.org
[6] https://zenodo.org
[7] https://guides.github.com/activities/citable-code/
[8] https://www.force11.org

exploring some of the more technical aspects of how to implement these recommendations (Katz et al., 2019)

In the last decade, the ASCL has enabled native software citation and software citation tracking in ADS and other indexers within astronomy and astrophysics; currently nearly 100 journals indexed by ADS show citations to software registered in the ASCL. The Asclepias project[9] – a collaboration between AAS publishing, ADS, and the Zenodo data archive – expands first-class support for software citation in AAS journals and further, rewards software authors who archive fixed versions of their software in Zenodo, thus ensuring future availability of these computational methods. While this project is currently scoped to AAS journals only, the changes being made to support the citation and indexing of software serve as an example for other journals to follow suit. We note that the Publications of the Astronomical Society of the Pacific (PASP) has published papers on astronomical software for several years and has been the "go to" place for software publication.

## 3.2 Strategies for elevating the role of software

Part of the challenge of elevating the role of software within academia is to establish actionable changes that improve the career prospects of those individuals writing and supporting research software and infrastructure tools. In this section, we outline several possible approaches.

***Software papers:*** One approach gaining traction across a number of research disciplines is to allow papers about software (i.e., without a stand alone research finding) to be published in "conventional" journals alongside other research papers, thereby making software more visible to the academic community, and giving software engineers a citable "creditable" entity (a paper) to include on their resume. Examples of journals within astronomy that demonstrate a willingness to follow this approach include PASP[10] and AAS publishing, which relatively recently changed its editorial policies to explicitly allow software papers in their publications[11]. More recently AAS publishing has announced a partnership with another journal (the Journal of Open Source Software, JOSS) specializing in software review (Vishniac & Lintott 2018).

***Improving support for software citation and indexing:*** Another key factor in raising the visibility of research software is to continue to improve software citation, citation training, and revealing these metrics to the world. As part of the work of the ASCL and the Asclepias project, software citations are counted in the astronomical literature and made visible on ADS. Further, ADS links papers to the cited software that enabled the reported research.

---

[9] http://adsabs.github.io/blog/asclepias
[10] https://iopscience.iop.org/journal/1538-3873
[11] https://journals.aas.org/policy-statement-on-software/

***Inform and educate the community about software contributions:*** Organizations play a critical role in improving the career prospects of those writing research software as they are responsible for hiring these individuals, evaluating their performance, and making decisions about possible promotions/career advancement. One immediately actionable approach is to encourage prospective employees and current staff to list software they have developed on their resumes and performance appraisals. This action would allow review committees to include software as part of their evaluations.

**Community prizes:** AAS has a collection of prizes for scientific merit, instrumentation, education, and service to the field. As it is an important part of scientific discovery, software contributions that have had a lasting positive impact on the field should also be recognized with a new dedicated prize and/or as a recognized example of merit within these other prize categories.

**Grants:** The amount of research funding secured is an established metric for evaluating an individual. As recommended in software the 2020 APC white paper entitled, *'Community Software For Astronomy in the 2020s: Challenges and Recommendations' (Tollerud et al.)*, allowing existing funding streams to be utilized for software development provides a simple mechanism for funding research software, but also signaling community recognition for the impact and relevance of the individual writing this software. Furthermore, widespread availability of grant funding in support of software development would provide a strong incentive for universities to hire technical astronomers into tenure track positions.

## 4. Conclusion

Software is already a critical part of modern research. The challenges of working with vast datasets produced by next-generation facilities will require ever more sophisticated software to analyze and interpret them thus making software, and software-creators, absolutely vital to the scientific success of these missions. While some progress has been made over the last decade to improve the career prospects of individuals devoting significant time to authoring software, there is still much work required to quantify the impact of research software and improve the career recognition of those individuals responsible for its development.

In this paper we have suggested seven actions that, if implemented, would provide mechanisms for measuring the impact of software in astronomy, give recognition to those individuals and teams making significant software contributions, and provide stable career paths within academia for these highly-skilled individuals.

The coming decade will see first light for community facilities such as LSST, DKIST, and WFIRST, each producing petabyte-scale datasets. In each case, scientific results from these facilities will rely upon a deep, broad collection of software components for data retrieval, analysis, and interpretation. Establishing reliable, transparent mechanisms for describing *what* software was used during an investigation is necessary for both awarding credit for this work, but also ensuring that these facilities result in reproducible, strong and verifiable science.